\documentclass[aps,prb,reprint,amsmath,amssymb,superscriptaddress,floatfix]{revtex4-2}
\usepackage{graphicx,xcolor,hyperref}
\usepackage{braket}
\usepackage[ignoreunlbld,norefs,nocites]{refcheck}
\begin{document}

\title{Resonating valence bond pairing energy in graphene by quantum Monte Carlo}
\author{S.\ Azadi}
\email{sam.azadi@manchester.ac.uk}
\affiliation{Department of Physics and Astronomy, University of Manchester, Oxford Road, Manchester M13 9PL, United Kingdom}
\author{A.\ Principi}
\affiliation{Department of Physics and Astronomy, University of Manchester, Oxford Road, Manchester M13 9PL, United Kingdom}
\author{T. D. K\"{u}hne}
\affiliation{Center for Advanced Systems Understanding, Untermarkt 20, D-02826 G\"orlitz, Germany}
\affiliation{Helmholtz Zentrum Dresden-Rossendorf, Bautzner Landstra{\ss}e 400, D-01328 Dresden, Germany}
\affiliation{TU Dresden, Institute of Artificial Intelligence, Chair of Computational System Sciences, N\"othnitzer Stra{\ss}e 46 D-01187 Dresden, Germany}
\author{M.\ S.\ Bahramy}
\affiliation{Department of Physics and Astronomy, University of Manchester, Oxford Road, Manchester M13 9PL, United Kingdom}
\date{\today}

\begin{abstract}
We determine the resonating-valence-bond (RVB) state in graphene using real-space quantum Monte Carlo with correlated variational wave functions. Variational and diffusion quantum Monte Carlo (DMC) calculations with Jastrow–Slater-determinant and Jastrow–antisymmetrized-geminal-power ans\"{a}tze are employed to evaluate the RVB pairing energy in graphene. Using a rectangular graphene sample that lacks $\pi/3$ rotational symmetry, we found that the single-particle energy gap near the Fermi level depends on the system size along the $x$-direction. The gap vanishes when the length satisfies $L_x = 3n\sqrt{3}d$, where $n$ is an integer and $d$ is the carbon–carbon bond length, otherwise, the system exhibits a finite gap. Our DMC results show no stable RVB pairing in this zero-gap case, whereas the opening of a finite gap near the Fermi level stabilizes the electron pairing. The DMC predicted absolute value of pairing energy at the thermodynamic limit for a finite-gap system is $\sim 0.48(1)$ mHa/atom. Our results reveal a geometry-driven electron pairing mechanism in the confined graphene nanostructures.
\end{abstract}

\maketitle

Many-body electron correlations in pristine graphene \cite{Novoselov2004,Geim2007,Castro2009,Novoselov2005}, are central to its most distinctive two-dimensional (2D) properties. The long-range Coulomb interaction is only marginally screened, leading to logarithmic renormalization of the Dirac Fermi velocity, breaking down the Fermi liquid picture, reduced quasiparticle weight, and scale-dependent effective coupling \cite{Kotov2012,Sarma2011,Cao2018}. The electrons near the Fermi level follow the Lorentz invariant theory, where the kinetic energy is a linear function of the momentum given by $K=v_F|\mathbf{p}|$ \cite{Castro2009} where $v_F$ is the Fermi-Dirac velocity and the property of a material. This linear behavior has a profound effect that causes the ratio of Coulomb to kinetic energy $\alpha$ to become independent of the electronic density, $\alpha\propto 1/\epsilon v_F$ where $\epsilon$ is also a material property. This is opposite to the Fermi liquid where the density parameter $r_s$ represents the weakly or strongly correlated regimes when it is small or large, respectively \cite{Azadi2024, Azadi2025}. Since the electronic density of states vanishes at the Dirac point, graphene is not a metal, but at the same time it is not an insulator either because it does not have a gap in the spectrum \cite{Novoselov2012,Guinea2010,NovoselovScience,Drut2009,Rotenberg2008,Sorella1992}. From an electron correlation point of view, the presence of the Dirac point makes the graphene a unique system. 

Another outcome of $K \propto p$ in graphene compared to most of systems with $K \propto p^2$ is related to the quantization of energy in confined electrons \cite{Sachdev1995}. If $K \propto p^2$ the quantized energy levels of electrons, which are confined to an area of size $a$, are proportion to space as $\Delta \varepsilon_0 \propto 1/a^2$, where as in graphene due to $K \propto p$, $\Delta \varepsilon \propto 1/a$. Hence, the size dependence of single particle energy states, which are described by plane waves and energy bands in periodic systems and determined by integration over the Brillouin zone, in small graphene sample size is different from those in the infinite system size limit. Hence in a nanoscale system size, crystal momentum is not a good quantum number, causing the bands touching at the Dirac point to become discrete levels. Also, since the Coulomb energy scales as $1/a$, the Coulomb effects are stronger in nanoscale graphene samples. Taken together, these unconventional behaviours imply that correlation-driven phenomena in graphene such as spin-liquid state \cite{Meng2010,Tran2011}, superconductivity\cite{Kotov2012}, and electron pairing in general, are strongly size-dependent. We emphasize that the coordinates of the Dirac point in reciprocal space are a repeating decimal, which has a crucial impact on single particle spectra as we discuss in this paper. Here we combine resonating valence bond (RVB) theory \cite{Anderson1987,Baskaran1988,Pauling,Glittum2024} with real-space quantum Monte Carlo methods to quantify and interpret the RVB pairing energy in graphene, and we clarify the role of the Dirac point in suppressing this energy and destabilizing the pairing. 

\begin{figure}
    \centering
    \includegraphics[width=0.45\linewidth]{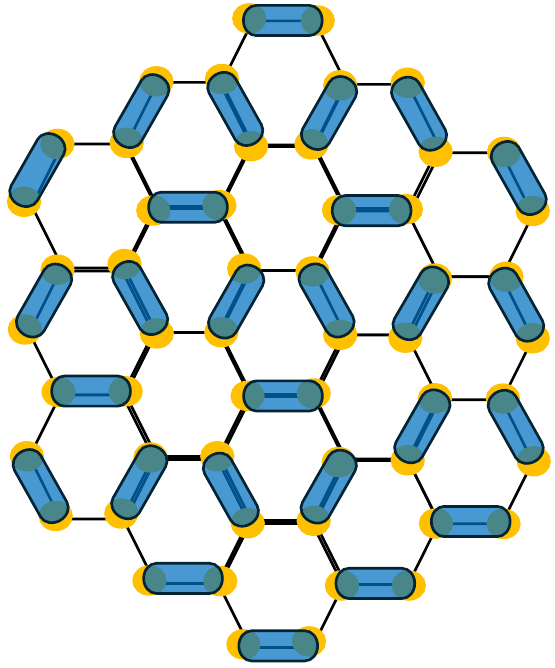}
    \caption{A spin-singlet dimer configuration (blue color) on graphene lattice. The  resonating valence bond (RVB) state is composed of a superposition of such configurations. Three $sp^2$ orbitals form the $\sigma$ band containing three localized electrons, while the bonding among the $p_z$ orbitals of different lattice sites generate a valence band ($\pi$ band) having one electron. The antibonding configuration generates the conduction $\pi^*$ band. }
    \label{fig:RVBConfig}
\end{figure}

 The RVB theory was originally proposed by Pauling to describe the electronic structure of aromatic molecules \cite{Pauling}. Within this framework, the bonding in graphene arises from a superposition of multiple valence bond configurations, reflecting the delocalized nature of its $\pi$ electrons (Fig.\ref{fig:RVBConfig}). Anderson applied the RVB concept to construct a nondegenerate ground state for a spin-half antiferromagnetic configuration on a triangular lattice\cite{Anderson1987,AndersonPRL87}. A valence bond is defined as a spin-singlet state given by $(S_i,S_j)=(\ket{S_{\uparrow_i}S_{\downarrow_j}}-\ket{S_{\downarrow_i}S_{\uparrow_j}})/\sqrt{2}$. An RVB state is a tensor product of the valence bond states:
\begin{equation}
    \ket{\Phi_{\text{RVB}}}=\sum_{i_1j_1\cdots i_nj_n} \alpha_{(i_1j_1\cdots i_nj_n)} \ket{(S_{i_1},S_{j_1})\cdots (S_{i_n},S_{j_n})}
\end{equation}
where $(S_{i_1},S_{j_1})\cdots (S_{i_n},S_{j_n})$ are singlet configurations over the lattice, and $\alpha_{(i_1j_1\cdots i_nj_n)}$ are variational parameters obtained by minimization of the ground state energy of a given Hamiltonian. The many-body wave function (WF) of the system is a summation over all possible configurations of dividing the lattice into dimers. The number of $\alpha_{(i_1j_1\cdots i_nj_n)}$ variational parameters is considerably large, causing difficulties for practical application of the above RVB WF. Noting that the Bardeen-Cooper-Schrieffer(BCS) conventional superconducting state is formed of spin-singlet Cooper pairs of electrons with opposite momentum, it was suggested that a practical RVB WF can be formed from BCS WF using a Gutzwiller projection $P_\infty$, $\ket{\Phi_{RVB}}  =P_{\infty}\ket{\Psi_{BCS}}$, in which $P_\infty=\Pi_i(n_{i,\uparrow} - n_{i,\downarrow})^2$, with $n_i$ denotes the particle density on site $i$, and $\ket{\Psi_{BCS}}=exp(\sum_{i,j}f_{i,j}c_{i,\uparrow}^{\dagger}c_{j,\downarrow}^{\dagger})\ket{0}$ where the pairing function in real space $f_{i,j}=f_{j,i}$ depends on relative distance $\mathbf{R}_i-\mathbf{R}_j$. In fact, $P_\infty$ eliminates all WF components with doubly occupied sites from the BCS WF and constrains the charge degrees of freedom. Therefore, the RVB WF is insulating with no density fluctuations. For a disordered antiferromagnetic system, it was suggested that the valence bond dimers in the RVB WF are dominated by short range pairs \cite{Anderson1987}, meaning $f_{i,j}\neq0$ only for nearest-neighbor (NN) sites, which results in a spin-liquid state without long-range spin order \cite{Zhou2017}. 

Quantum Monte Carlo calculations of correlated electrons on a graphene lattice predict a short-ranged RVB quantum spin-liquid state, which proposes the possibility of unconventional superconductivity via doping with $d+id$ pairing symmetry\cite{Meng2010,Tran2011,Pathak2010}. The stability of a spin-liquid state in the Hubbard model on the graphene lattice close to an antiferromagnetic Mott insulator\cite{Meng2010} could be similar to some organic antiferromagnet with a possible spin-liquid ground state \cite{Shimizu2003,Norman2016}. The application of RVB theory is not limited to the strongly correlated regime near the Mott transition or the spin-liquid model, but it can be expanded to describe the electronic structure and pairing mechanism in strongly correlated real systems \cite{Casula2013, BeccaSorella}. 

The RVB WF used in this work is defined as $\ket{\Phi_{RVB}}$=J$\ket{\Psi_{AGP}}$ where J is the Jastrow term responsible for the short range Coulomb interaction and $\ket{\Psi_{AGP}}$ is an antisymmetrized geminal power (AGP) determinant part given by
\begin{equation}
    \ket{\Psi_{AGP}(\mathbf{R})}=\mathcal{A} \Pi_{i=1}^{N_{el\downarrow}} \phi(\mathbf{r}_{i}^{\uparrow}, \mathbf{r}_{i}^{\downarrow} )  
\end{equation}
where $\mathcal{A}$, $\mathbf{R} = \left \{ \mathbf{r}_{1}^{\uparrow}, \cdots, \mathbf{r}_{N_{el\uparrow}}^{\uparrow}, \mathbf{r}_{1}^{\downarrow},\cdots, \mathbf{r}_{N_{el\downarrow}}^{\downarrow} \right \} $, and $\phi(\mathbf{r}_{i}^{\uparrow}, \mathbf{r}_{i}^{\downarrow}) = \phi(\mathbf{r}_{i}^{\downarrow}, \mathbf{r}_{i}^{\uparrow} )$, are the antisymmetrization operator, the $3N_{el}$-dimensional vector of electron coordinates, and a symmetric orbital function describing the singlet pairs, respectively. We used molecular orbitals (MOs) basis to expand the pairing function
\begin{equation}
 \phi(\mathbf{r}^{\uparrow}, \mathbf{r}^{\downarrow}) = \sum_{i=1}^{M} \alpha_i \phi_i^{MO}(\mathbf{r}^{\uparrow}) \phi_i^{MO}(\mathbf{r}^{\downarrow}),
\end{equation}
where the sum is over $M\geq N_{el}/2$ MOs that were expanded in a Gaussian one-particle basis set ${\chi}$ centered on the atomic position $\phi_i^{MO}(\mathbf{r}) = \sum_j \beta_{ij} \chi_{j}(\mathbf{r})$ \cite{Sorella2015}. The variational freedom of the WF increases when $M>N_{el}/2$ with respect to the minimal case of $M=N_{el}/2$ which represents the Jastrow Slater Determinant (JSD) WF \cite{BeccaSorella}. A specific choice of $M$ is important for enhancing the JAGP WF accuracy to capture the pairing energy. In our simulations $M$ is the minimum number of MOs used to determine a product of independent HF WFs of single atoms \cite{Marchi2009,Marchi2011,OzoneTurboRVB,Casula2003,Casula2004}. An uncontracted Gaussian basis of $8s6p4d$ orbitals was used for the carbon atom presented by correlation consistent pseudopotentials with four valence electrons \cite{ccECP1,ccECP2}.  The MOs were obtained by solving the Kohn-Sham equations on a grid \cite{Azadi2010} with the local density approximation (LDA) \cite{lda}. The advantage of using AGP WF, which is the electron-conserving version of the BCS WF, is that it includes a single determinant with static correlation. 

The dynamic correlation is included in the Jastrow term, which plays a role in describing a spin-liquid state and is defined by a weight factor $J(\mathbf{R})=exp(\sum_{i<j} u(\mathbf{r}_i, \mathbf{r}_j))$ where $\mathbf{R}$ is the 3N-dimensional configuration of the electron positions $\mathbf{r}_i$. The Jastrow correlation is a positive decaying function of the electron distance $r=|\mathbf{r}_i-\mathbf{r}_j|$ and is given by $u(r)=r/(2(1+ar))$, with $a$ as a variational parameter\cite{Fahy90}.  For Jastrow single-particle orbitals, we used uncontracted Gaussian basis sets of $4s3p$. All variational parameters in trial WFs were optimized by energy minimization using two techniques of linear basis \cite{Umrigar2007} and stochastic reconfiguration (SR) \cite{Sorella98}. 
\begin{figure}[!htb]
    \centering
    \begin{tabular}{cc}
          \includegraphics[width=0.5\linewidth]{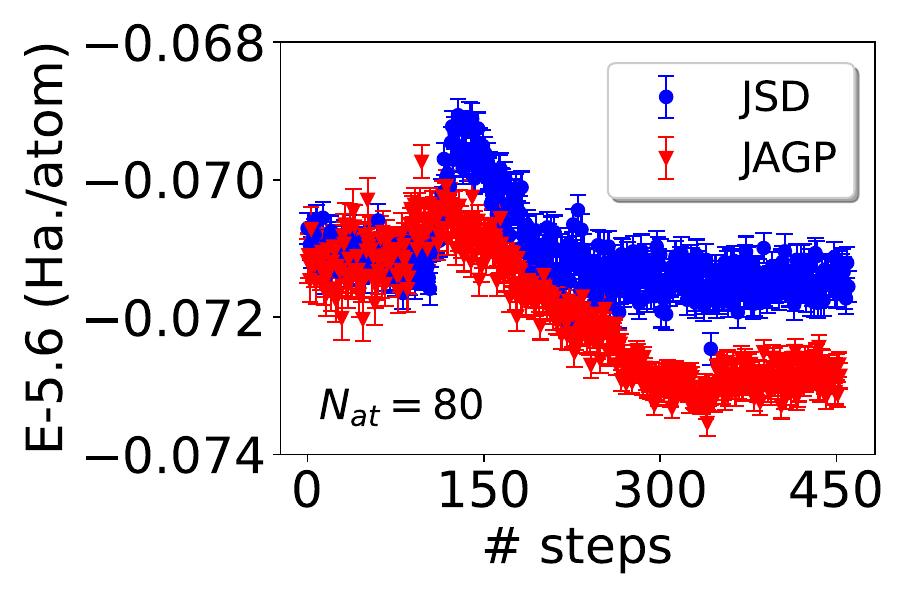} &
          \includegraphics[width=0.5\linewidth]{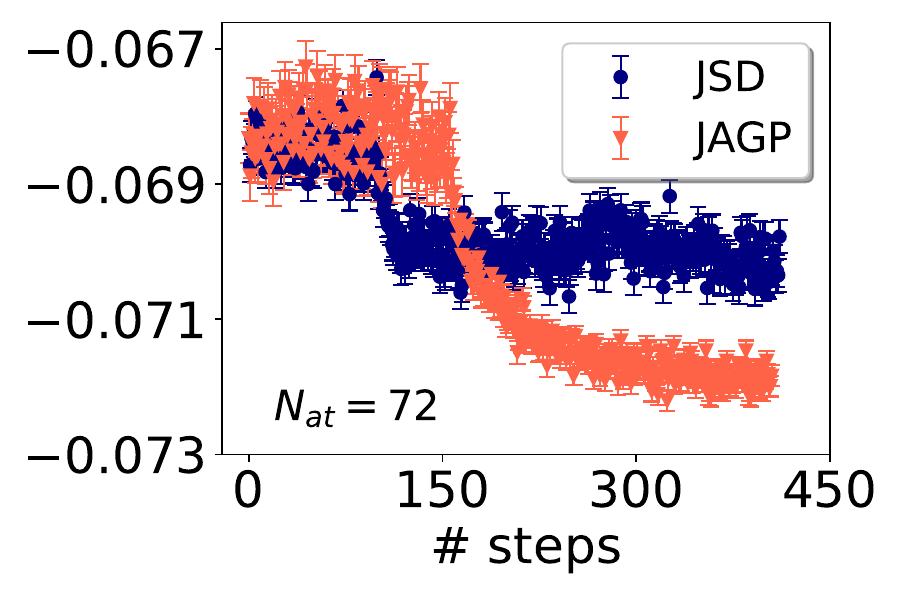} \\
    \end{tabular}
    \caption{VMC energy as a function of the number of optimization steps computed using JSD and JAGP wave functions for finite-gap system with number of atoms in simulation cell $N_{at}=80$ (left panel) and zero-gap system with $N_{at}=72$ (right panel). }
    \label{fig:E_opt}
\end{figure}

We compare total energies computed using variational quantum Monte Carlo (VMC) \cite{Matthew2001,BeccaSorella} and diffusion quantum Monte Carlo (DMC) \cite{Matthew2001,Ceperley78,Ceperley80}, two well-established stochastic methods for many-body electronic structure calculations. In the VMC approach, the total energy is calculated as the expectation value of the Hamiltonian with respect to a chosen trial WF, which is optimized to minimize the variational energy. DMC improves upon this by using a stochastic projection method to evolve the trial WF in imaginary time, effectively filtering out excited-state components and projecting onto the ground-state wave function, within the fixed-node approximation \cite{Melton2017}. All of our VMC and DMC results are obtained using two classes of trial WFs of JAGP and JSD. To quantify the strength of electron pairing in the system, we define the pairing energy $\delta_P$ as the difference between the total energies obtained with the JSD and JAGP WFs $\delta_P = E_{\text{JAGP}} - E_{\text{JSD}}$. A negative value of $\delta_P$ indicates that the JAGP WF captures the effective attraction between opposite spin electrons. If $\delta_P < 0$ in the thermodynamic limit, it suggests that the BCS condensation of the electron singlets is stable. More details on the WFs used and the technical implementation are provided in the previous works \cite{Marchi2009,Marchi2011,Casula2003,Azadi25,TurboRVB,BeccaSorella}.

We construct a rectangular primitive cell consisting of four carbon atoms, defined by the lattice vectors $\mathbf{a} = \sqrt{3}d~\mathbf{\hat{x}}$ and $\mathbf{b} = 3d~\mathbf{\hat{y}}$, where $d = 1.42~\AA$ denotes the nearest-neighbor (NN) C distance \cite{suppl}. This rectangular cell is twice the size of the conventional hexagonal graphene primitive cell, which contains two atoms and has lattice vectors $\mathbf{a} = d~\mathbf{\hat{x}}$ and $\mathbf{b} = -\frac{d}{2}~\mathbf{\hat{x}} + \frac{\sqrt{3}d}{2}~\mathbf{\hat{y}}$. Importantly, the $\mathbf{a}$ directions of both cells are aligned, which allows a straightforward comparison. Direct comparison between 4-atom rectangular and 2-atom hexagonal unit cells and their Brillouin zones (BZ) are presented in the supplementary materials \cite{suppl}. 

The simulation supercells are constructed by tiling the rectangular primitive cell into $n \times m$ arrays, where $n$ and $m$ are integers satisfying $2 \leq n, m \leq 6$, resulting in simulation cells with atom counts ranging from 16 to 80. These rectangular supercells do not preserve the sixfold rotational ($\pi/3$) symmetry of the infinite graphene lattice. As a result, they explicitly break rotational symmetries of certain superconducting order parameters, such as real-valued $d_{xy}$ and $d_{x^2-y^2}$ pairings. In contrast, the energetically favorable chiral $d + id$ pairing symmetry in the ground state wave function retains the full symmetry of the lattice \cite{Schaffer2007}. We restrict ourselves to real-valued wave functions obtained at the $\Gamma$-point, which are appropriate for time-reversal symmetric systems without intrinsic chiral or magnetic order\cite{Mitas2025}. To determine the pairing energy in the thermodynamic limit, we perform a linear extrapolation of the pairing energy estimator $\delta_p$ as a function of inverse system size ($1/N$), where $N$ is the number of atoms in the simulation cell.

We found that WF optimization plays a critical role in accurately capturing the pairing energy. We began by systematically optimizing the JSD WF, first optimizing the Jastrow coefficients and two-body correlation terms, followed by the optimization of the Jastrow Gaussian exponents. At this stage, we obtained the standard form of the WF commonly used in QMC simulations, where the Jastrow factor is fully optimized and the Slater determinant is taken from a mean-field approach, DFT in our case. We then proceeded with a full energy optimization of both the Jastrow and Slater determinant components using the linear method \cite{Umrigar2007}, followed by an improvement with the SR technique \cite{Sorella98}. The resulting energy of the JSD wave function is shown in Fig.~\ref{fig:E_opt}. The optimized exponents of Jastrow and determinant parts of the JSD WF were then used to build and convert the JAGP WF where the number of MOs is increased to represent a product of independent HF WFs of the C atom \cite{Marchi2009,Marchi2011}. The additional MOs were reinitialized by performing an additional DFT simulation. We then reoptimized all the variational parameters by taking the above steps to determine the JAGP WF. We carried out all these steps for the simulation cells used in this work and compared the JAGP energy with JSD. Out of thousands of optimization iterations, we show the last few hundred steps in which the JAGP and JSD energies are competitive \cite{suppl}. Figure.~\ref{fig:E_opt} illustrates the VMC energy as a function of the number of optimization steps obtained using JSD and JAGP WFs for two system sizes of $N_{at}=80$ and $N_{at}=72$ atoms in the simulation cell. The single-particle DFT band structure calculations \cite{suppl} yield a finite gap with value of $E_g=2.05$ eV for $N_{at}=80$ and zero gap for $N_{at}=72$ systems. Both plots demonstrate the energy gain by optimizing the AGP pairing determinant. As it is discussed in the following, although the VMC-JAGP is lower than VMC-JSD the DMC energies predict a different trend. 
\begin{figure}
    \centering
          \includegraphics[width=0.75\linewidth]{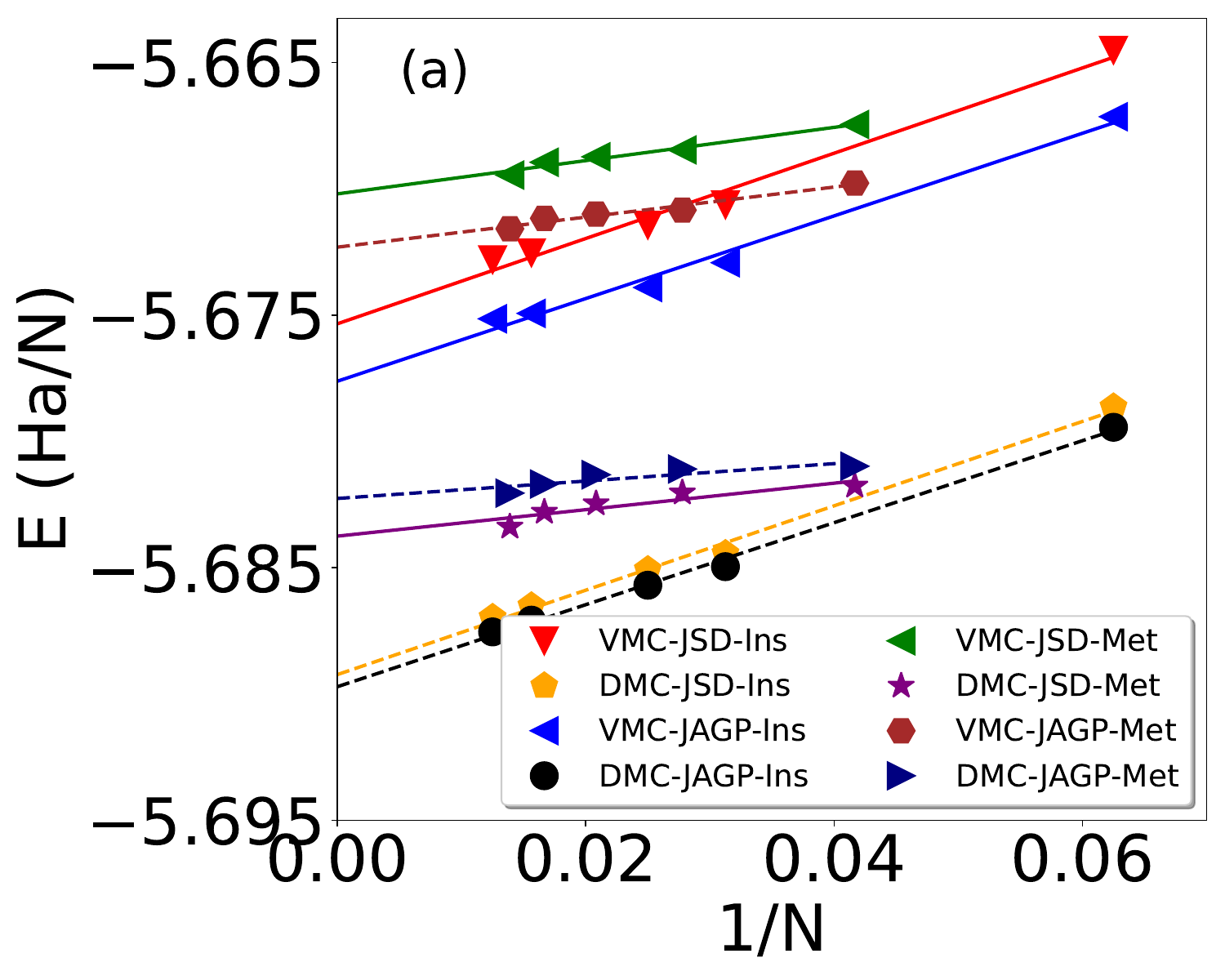} \\
          \includegraphics[width=0.75\linewidth]{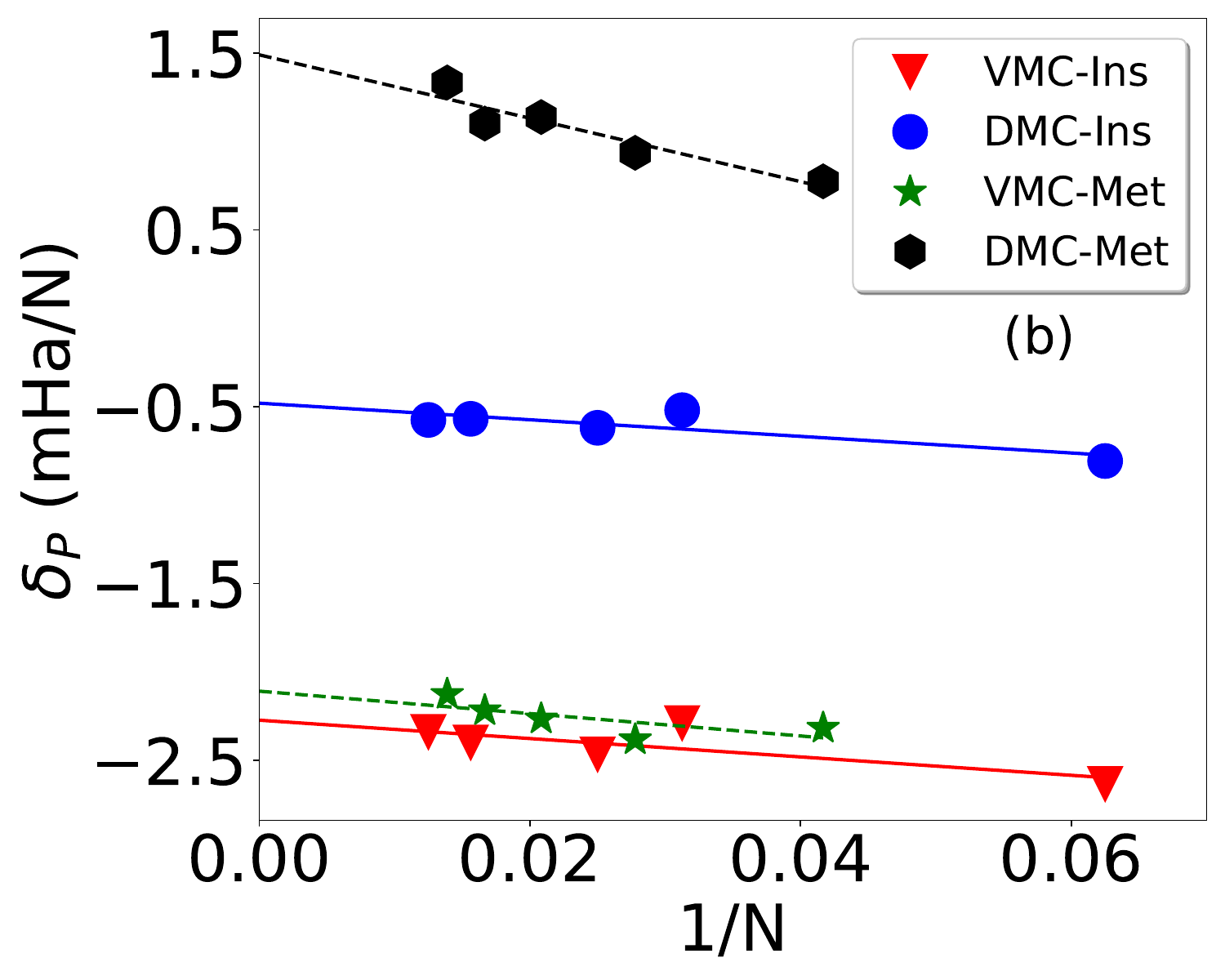} \\
    \caption{(a) VMC and DMC energies for finite gap (Ins) and zero gap (Met) systems obtained with JSD and JAGP WFs as a function of system size. (b) VMC and DMC pairing energies for finite gap (Ins) and zero gap (Met) systems obtained with JSD and JAGP WFs as a function of system size. The error bars are smaller than data points size. }
    \label{fig:E_pair}
\end{figure}

The VMC and DMC energies in the thermodynamic limit were obtained by linearly extrapolating the QMC energies as a function of $1/N$, where $N$ is the number of atoms in the simulation cell (Fig.~\ref{fig:E_pair}). Calculations were performed using both JSD and JAGP WFs for systems with finite and vanishing energy gaps. Our results show that, in the thermodynamic limit, the VMC and DMC energies of the finite-gap $E_g\neq0$ system are lower than those of the zero-gap $E_g=0$ system for both types of WFs, indicating enhanced stability of the gapped phase. Fig.~\ref{fig:E_pair} (b) shows the pairing energy $\delta_P$ for $E_g\neq0$ and $E_g=0$ systems as a function of system size. The VMC-$|\delta_P|$ at the thermodynamic limit is larger than DMC for both systems with $E_g\neq0$ and $E_g=0$. The DMC-$\delta_P$ at the thermodynamic limit, which is more accurate than VMC, clearly show no stability of RVB pairing for $E_g=0$ system. Our DMC results do not show any energetic stabilization of the JAGP WF relative to JSD for the simulation cell size with the lattice parameter along the x-axis $a^*=3na_0$ with $n=1, 2$ and $a_0$ the primitive lattice parameter. For structures characterized by the lattice parameter $a^*$, the magnitude of the pairing measure $\lvert\delta_P\rvert$ obtained from both VMC and DMC is smaller than for the other geometries examined. Moreover, DMC yields $\delta_P>0$ for the $a^*$ geometries, indicating the absence of RVB-driven pairing.  

\begin{figure}
    \centering
    \begin{tabular}{c c}
    \includegraphics[width=0.45\linewidth]{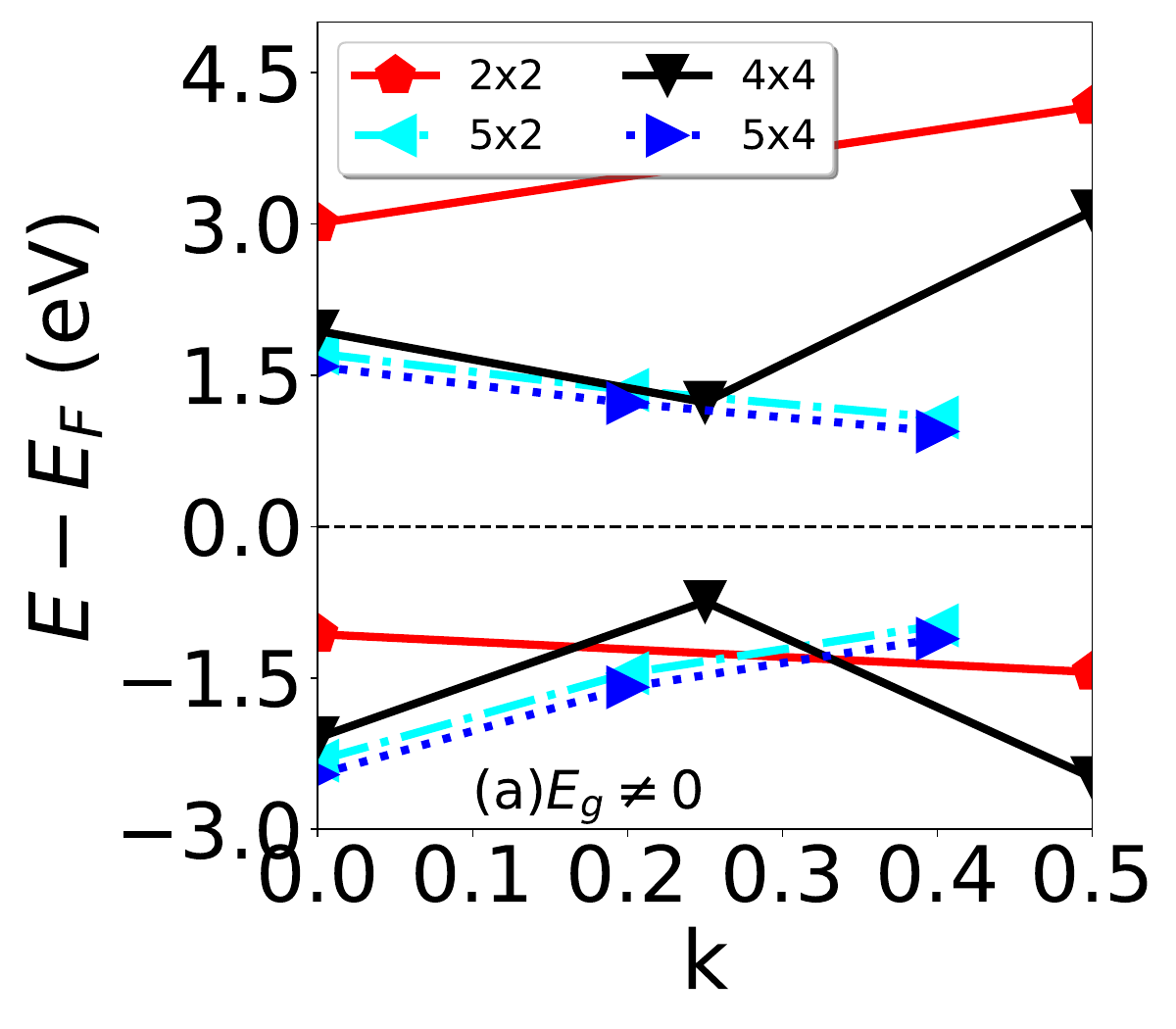}&
    \includegraphics[width=0.45\linewidth]{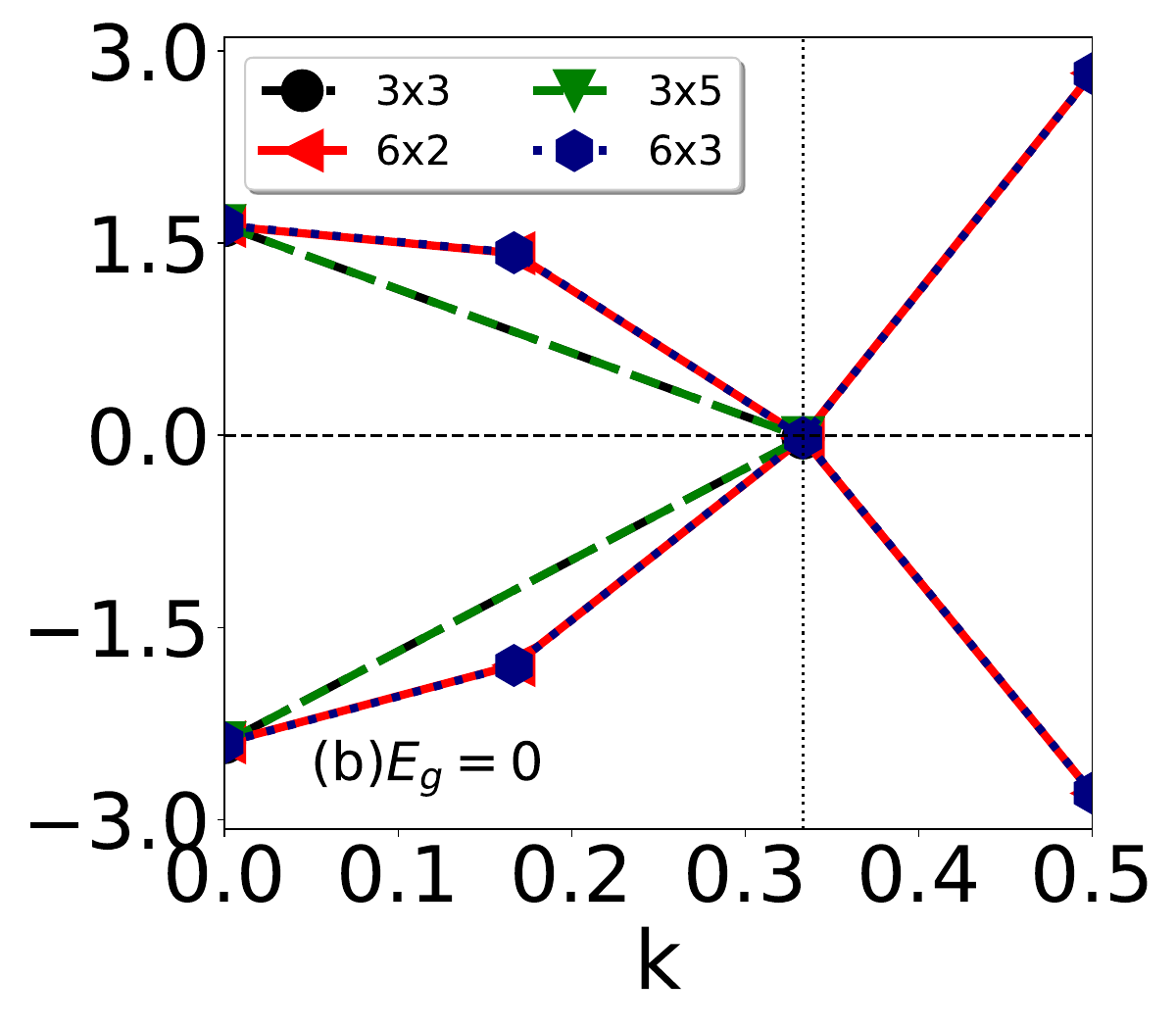}\\
    \includegraphics[width=0.45\linewidth]{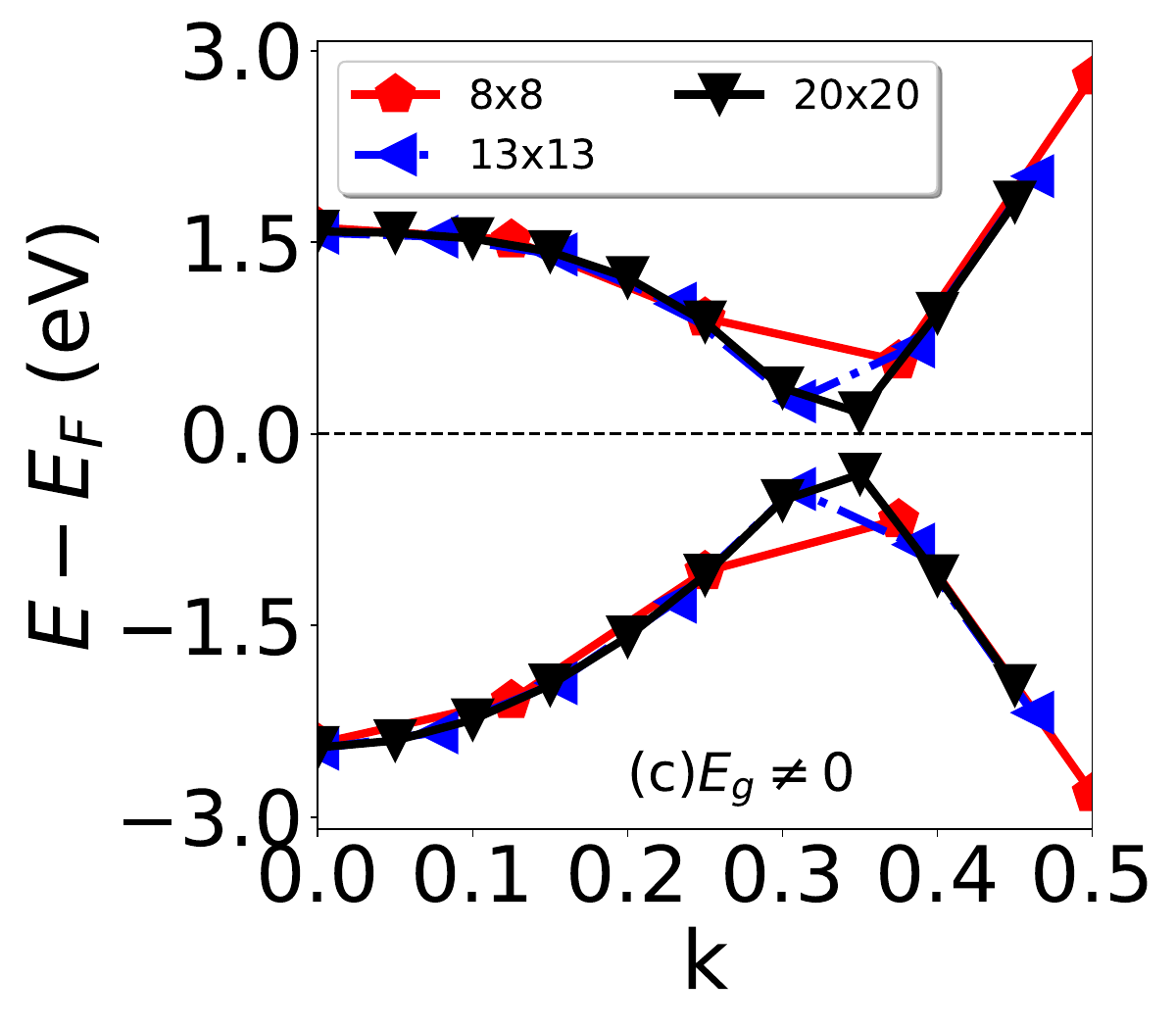}&
    \includegraphics[width=0.45\linewidth]{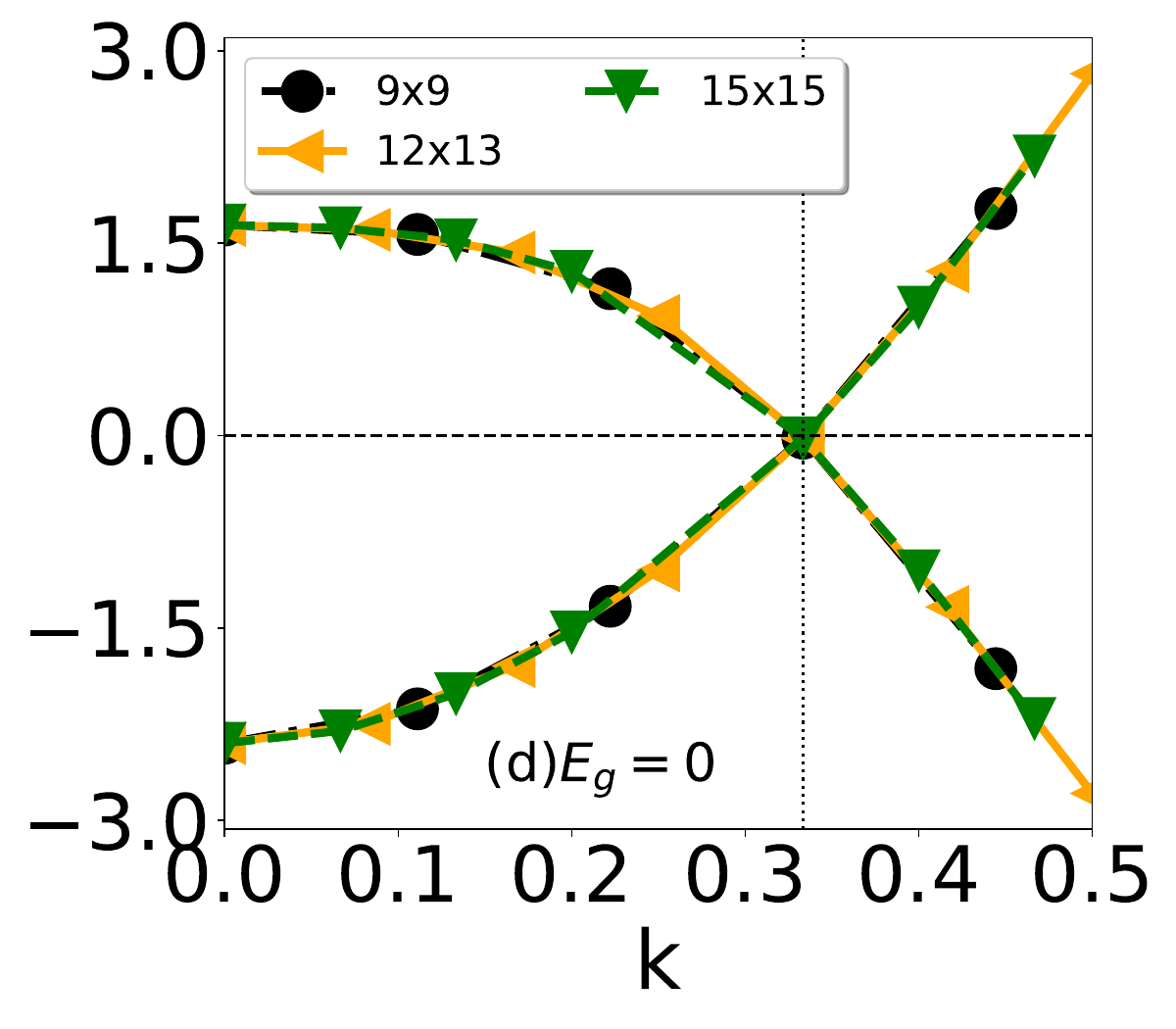}\\
    \includegraphics[width=0.45\linewidth]{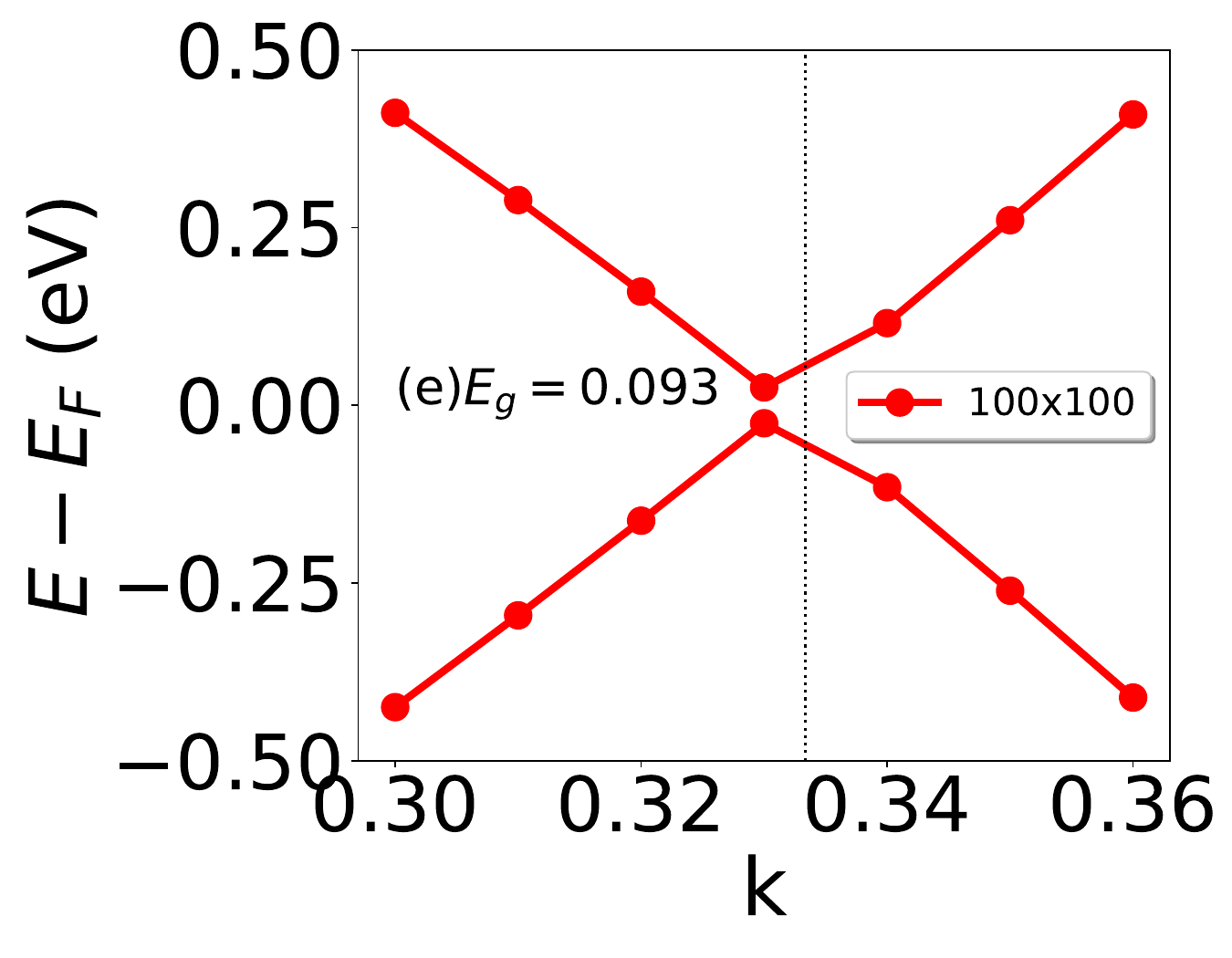}&
    \includegraphics[width=0.45\linewidth]{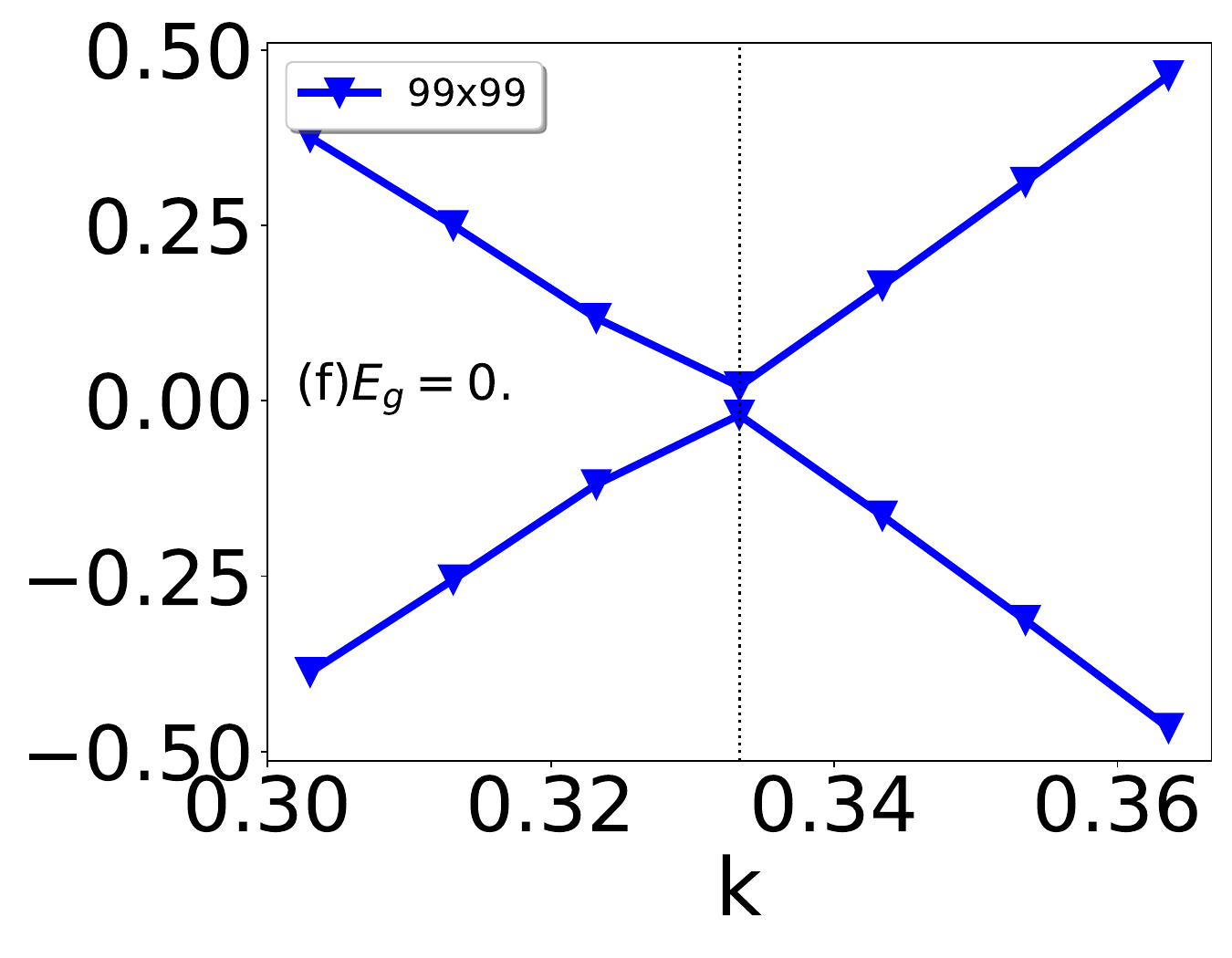}\\    
    \end{tabular}
    \caption{Single particle energy band as a function of momentum computed by DFT. The valence and conduction band along $\Gamma(0,0)\rightarrow X(\frac{1}{2},0)$ obtained for different $\mathbf{k}$-meshes. The horizontal and vertical lines at $E=0$ and $k=\frac{1}{3}$ represent the Fermi energy and Dirac point, respectively. (a) and (b) show bands for non-zero $E_g\neq0$ and zero gap $E_g=0$ systems for which the RVB pairing energy is obtained by VMC and DMC. (c) and (d) show bands for larger system sizes with non-zero and zero gaps. Comparing the bands obtained by $\mathbf{k}$-mesh $13\times13$ and $12\times13$ demonstrate that vanishing gap is only depend on $k_x$. (e) and (f) show bands near the Dirac point obtained using $\mathbf{k}$-meshes of $100\times100$ with $E_g=0.093$ eV and $99\times99$ with $E_g=0$. }
    \label{fig:Bands}
\end{figure}

We found that the geometry dependent behavior of $\delta_P$ is due to the occupation pattern of single-particle energy state and momentum quantization due to PBC in the BZ. In the $a^*=3na_0$ geometries, the folded $\mathbf{k}$-vectors contains the $\mathbf{k}=(\tfrac{1}{3},0,0)$ point in primitive reciprocal-lattice units, at which valence and conduction bands touch on the Fermi level. The DFT occupations spectra \cite{suppl} reveal that only at this $\mathbf{k}$ point the degenerate bands just below and above $E_F$ carry $\sim0.5$ occupancy, signaling a metallic configuration. In contrast, for the other studied geometries the energy band occupations at every $\mathbf{k}$ point are integer, 1 below $E_F$ and 0 above, resulting in insulating phase. To investigate this further, we computed DFT band structures for a series of $\mathbf{k}$-point meshes, grouping them by whether the number of divisions along $k_x$ was a multiple of three $N_x=3n$ or not $N_x\neq3n$, with integer $n$ (Fig. \ref{fig:Bands}). We find that meshes with $N_x\neq3n$ exhibit a finite gap near the Fermi level, whereas $N_x=3n$ yields a vanishing gap at the Dirac point. This observation is independent of $k_y$, as it can be observed from band structures in Fig.~\ref{fig:Bands}. 

The ground state DMC energy depends on the nodal surface of the trial WF. The nodal surface of the WFs used in our DMC simulations is determined by the occupied subspace of single-particle orbitals and any unitary rotation within that subspace leaves the node unchanged. Including the Jastrow correlation function in the WF does not also affect the nodal surface. Appearance a gap near the Fermi energy in one-particle spectra separates the occupied subspace from the unoccupied. The degenerate occupied Dirac-cone states on the Fermi energy changes the nodal surface of the WF and consequently introduces a difference between DMC-JSD and DMC-JAGP energies.  The key point is that a vanishing gap and strictly linear single-particle dispersion $E(k)$ occur only when the $k$-mesh samples the Dirac point exactly meaning that the Brillouin-zone coordinates containing recurring decimal $1/3$ fractions. For meshes with $N_x\neq3n$ in reciprocal space, or equivalently, real-space periodicities with $a^*\neq3n\,a_0$, the Dirac point is not sampled, the ideal linear $E(k)$ is perturbed, and a gap opens near the Fermi level. Increasing $N_x\neq3n$ in reciprocal space or system size $a^*\neq3n\,a_0$ reduces the gap which never reaches to absolute zero but $\epsilon>0$. This can be observed from Figs.\ref{fig:Bands} (e) and (f) showing the bands near Dirac point for $\mathbf{k}$-meshes of $100\times100$, with $\sim 0.1$ eV gap, and $99\times99$ with zero gap, respectively. This has to be noted that the value of gap is predicted by DFT which is well-know in underestimating the energy gap near the Fermi energy. The exact value of the band gap and its nature need to be determined by many-body based methods. 

Our results indicate that selecting certain longitudinal dimensions for a rectangular nanoscale graphene sheet opens a finite, size-dependent energy gap and, in turn, yields an attractive RVB pairing energy. In this regime, the finite cell’s BZ does not sample the Dirac point exactly. Instead it captures nearby k-points where the gap appears. To illustrate, we performed DFT simulations for two system sizes with 676, with lattice parameters a=3.197 and b=5.538 nm,  and 624, with lattice parameters a=2.951 and b=5.538 nm, C atoms corresponding to $\mathbf{k}$-meshes of $13\times13$ and $12\times13$, respectively, with band structures shown in Fig.~\ref{fig:Bands} (c) and (d). DFT further indicates that the 676-atom system size is insulating with a gap of $E_g=0.69$ eV, whereas the 624-atom system size remains gapless with a degeneracy at the Fermi level \cite{suppl}. Guided by our QMC results, we therefore expect RVB pairing to be stabilized in the 676-atom system, consistent with the emergence of a superconducting state. Appearing a small gap in a graphene sample reduces low-energy density of states (DOS) and therefore disfavoring weak-coupling pairing and consequently changes the balance between $s$ or $d$ wave states and screening. Our results suggest that the superconductivity in geometrical driven finite gap graphene sheet generated by electron correlation through RVB mechanism instead of weak electron-phonon coupling. Breaking the linear behavior of $E(k)$ and opening a gap near the Fermi energy can introduce other properties which require further studies. For instance, a small gap replaces massless, chiral Dirac fermions by massive ones, turning on Berry curvature and valley physics, and possibly suppressing Klein tunneling \cite{Kotov2012,Sarma2011} which are depend on the nature of the gap. 

In summary, we performed VMC and DMC simulations using two types of WFs, JSD and JAGP, to calculate the RVB pairing energy $\delta_P$ in graphene. The QMC energies were computed for finite system sizes using simulation cells constructed by tiling a four-atom rectangular primitive cell into $n \times m$ arrays, with $n$ and $m$ being integers. We found that the system size along the $x$-axis, $L_x$, plays a crucial role in determining the electronic structure and correlation effects. When $L_x = 3n\sqrt{3}d$, where $n$ is an integer and $d = 1.42$~\AA\ is the C–C bond length, the system exhibits a vanishing one-particle energy gap $E_g = 0$. In contrast, for $L_x \ne 3n\sqrt{3}d$, the system develops a finite energy gap $E_g \ne 0$ near the Fermi level. Based on the extrapolated VMC and DMC energies in the thermodynamic limit, the gapped $E_g \ne 0$ system is more stable than the gapless $E_g = 0$ one, suggesting a preference for a non-metallic ground state. Furthermore, the DMC-$\delta_P$ shows no stability in the metallic systems, whereas it reaches a value of approximately $0.48(1)$ mHa/atom in the thermodynamic limit for the insulating case.

We acknowledge the support of the Leverhulme Trust under the grant agreement RPG-2023-253. S. Azadi and T.D. K\"{u}hne acknowledge the computing time provided to them on the high-performance computers Noctua2 at the NHR Center in Paderborn (PC2).

\bibliography{main}

\end{document}